\documentclass[prd,twocolumn]{revtex4}
\usepackage{graphicx}

\begin{document}

\title {\large Fermionic partner of Quintessence field as candidate for dark
    matter }

\author{ Xiao-Jun Bi }
\affiliation{ Institute of High Energy Physics, Chinese Academy of 
Sciences, P.O. Box 918-4, Beijing 100039, People's Republic of China}

\author{ Mingzhe Li }
\affiliation{ Institute of High Energy Physics, Chinese Academy of 
Sciences, P.O. Box 918-4, Beijing 100039, People's Republic of China}

\author{ Xinmin Zhang }
\affiliation{ Institute of High Energy Physics, Chinese Academy of 
Sciences, P.O. Box 918-4, Beijing 100039, People's Republic of China}

\date{\today}

\begin{abstract}

Quintessence is a possible candidate for dark energy.
In this paper we study the phenomenologies of
the fermionic partner of Quintessence, the Quintessino. Our results 
show that, for suitable choices of the model parameters, the Quintessino is 
a good candidate for cold or warm dark matter. In our scenario, dark
energy and dark matter of the Universe are connected 
in one chiral superfield. 

\end{abstract}

\maketitle

Recent data from type Ia supernovae \cite{pel} and cosmic
microwave background (CMB) radiation \cite{wmap} have provided strong 
evidences for a spatially flat and
accelerated expanding universe at the present time.
 In the context of
Friedmann-Robertson-Walker cosmology, this acceleration is attributed to 
the domination of a component, dubbed dark energy \cite{tur}.
The simplest candidate for dark energy
seems to be a remnant small cosmological constant. However, 
many physicists are attracted by
the idea that dark energy is due to a dynamical component, such as a 
canonical scalar field $Q$,
named {\it Quintessence} \cite{rp}. 
The data from WMAP\cite{wmap} indicate that 
the potential of the Quintessence field 
around present epoch should be very flat. Consequently its
effective mass will be extremely small, $m_Q\leq
H_0\sim 10^{-33}$ eV.

When coupled to the ordinary matter,
Quintessence boson will induce
a long range force and cause the physical constant vary with time. The current 
experiments have put strong limits on these couplings. 
In Ref.\cite{carroll}, Carroll considered a 
possibility of suppressing such couplings by imposing 
an approximate global shift symmetry, 
$Q \rightarrow Q + {\rm C}$, where $C$ is a constant. In 
this case, Quintessence behaves like a Pseudo-Goldstone boson. 
With this approximate symmetry, Carroll \cite{carroll} further 
explicitly proposed an interaction of the form
\begin{equation}
\label{qphoton}
\mathcal{L}_{Q\gamma\gamma}=\frac{c}{M_{pl}}QF_{\mu\nu}\tilde{F}^{\mu\nu}\ ,
\end{equation}
where $F_{\mu\nu}$ is the
electromagnetic field strength tensor and $\tilde{F}_{\mu\nu}$
is its dual. 
The current data do not strongly constrain the coefficient 
$c$, $c \leq 3 \times 10^{-2}$, which interestingly opens 
a possibility to detect the effects of Quintessence 
on the rotation of the plane of polarization of light
coming from distant sources in the near future. 

From the point of view of particle physics, fundamental interactions,
as widely believed, may be supersymmetric (SUSY)
beyond the TeV scale. In a SUSY theory, the 
Quintessence boson will be accompanied by a 2-component neutral fermion 
($\tilde{Q}$), {\it Quintessino}, and a scalar ($\sigma_q$), the 
Squintesson. The Quintessence, as argued above, is taken as a 
pseudo-Goldstone boson, so one would expect
its fermionic superpartners also light. A naive 
dimensional analysis in a model 
independent way indicates $m _{\tilde{Q}} 
\sim \mathcal{O}(M_{SUSY}^2/ \Lambda )$, where $\Lambda$ corresponds to 
the decay constant of the Pseudo-Goldstone boson, the Quintessence here.
It may be possible, however, in the similar way as for
axino\cite{axino} and Majorino\cite{majo} that the Quintessino
receives a large mass in a specific model. In this paper we will 
study the phenomenologies of Quintessino and, for our discussions,
we take its mass as a free parameter. 

In the minimal supersymmetric standard model (MSSM) with a conserved 
R-parity, the lightest SUSY particle (LSP), taken usually as the lightest
neutralino, $\chi^0_1$, is stable and serves as
an ideal candidate for cold dark matter (CDM).
However, if the Quintessino is lighter than $\chi^0_1$, 
the interaction in Eq. (\ref{qphoton}) with $c \sim 10^{-2}$
leads the $\chi^0_1$ decay away so that the neutralino can not serve as 
CDM, unless the coefficient $c$ is smaller than
$c < 10^{-6}$ and the neutralino is actually stable.
We further study the possibility that the $\chi^0_1$ decay product,
the Quintessino, forms dark matter of the Universe. 
However, the released electromagnetic energy from the neutralino decay 
will be too much to be compatible with the Big Bang nucleosynthesis 
(BBN)\cite{bbn} and CMB \cite{cmb} data.  
To resolve the conundrum, we introduce new couplings between 
the Quintessence and the matter fields.  We will show that
the Quintessino can serve as a good candidate for cold or warm dark 
matter for suitable choices of the parameters.
In our model, the dark energy, Quintessence, and dark matter, Quintessino,
are unified in one chiral superfield, similar to the quark and lepton in
the same representation of a gauge group in the unified theory.

We start our discussions with supersymmetrizing Eq. (\ref{qphoton}).
The relevant part of the Lagrangian responsible for the neutralino decay 
 is given by
\begin{equation}
\label{qptino}
\mathcal{L}_{\tilde{Q}\tilde{\gamma}\gamma}
=\frac{c}{M_{pl}}\bar{\tilde{Q}}\gamma^5\sigma^{\mu\nu}
\tilde{\gamma}F_{\mu\nu}\ .
\end{equation}
Taking the neutralino to be Bino-like, a simple calculation gives its
decay width 
\begin{equation}
\Gamma(\tilde{B}\to \tilde{Q}\gamma)=
\frac{1}{2\pi}(\frac{c}{M_{pl}})^2 \cos^2\theta_W m_{\tilde{B}}^3
( 1-x^2 )^3 \ ,
\end{equation}
with $x=\frac{m_{\tilde{Q}}}{m_{\tilde{B}}}$.
Taking the coefficient $c$ to be $3\times 10^{-2}$,
$m_{\tilde{B}}$ around the electroweak scale and 
assuming $m_{\tilde{Q}}$
smaller than $m_{\tilde{B}}$, the lifetime of the Bino will be
in the range of $10^7 - 10^{12}$sec, which is much 
shorter than the age of the Universe. 
To stabilize the Bino, so that it forms the CDM of the
Universe, the coupling constant $c$ has to be smaller than $10^{-6}$.
In this case, however, 
it may become impossible to detect the Quintessence effect
in the polarization studies, as suggested
by Carroll.

\begin{figure}
\includegraphics[scale=0.35]{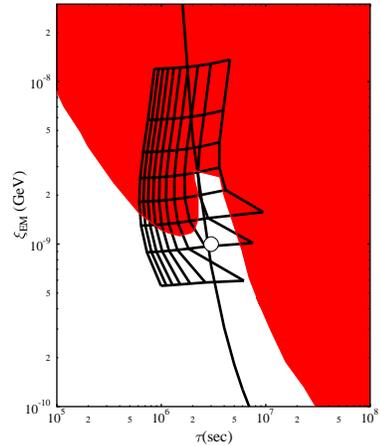}
\caption{\label{bino}
Lifetime $\tau$ and energy release $\xi_{EM}$ in Bino decay 
for $m_{\tilde{B}}=300 GeV, 400 GeV, \ldots, 1.2 TeV$ (from right
to left) and $x_Q=0.1, 0.2, \ldots, 0.8$ (from top to bottom).
We take $c=2\pi$ and $m_{h^0}=115 GeV$. 
The contour of the chemical potential for photon 
distribution function is $\mu=9\times 10^{-5}$,
with the region to the right of it being excluded by CMB\cite{pdg}.
The shaded regions are excluded by BBN data\cite{bbn}.
The circle represents the best fit region with
$(\tau, \xi_{EM})= (3\times 10^6 sec, 10^{-9} GeV)$\cite{bbn}.
}
\end{figure}

\begin{figure}
\includegraphics[scale=0.25]{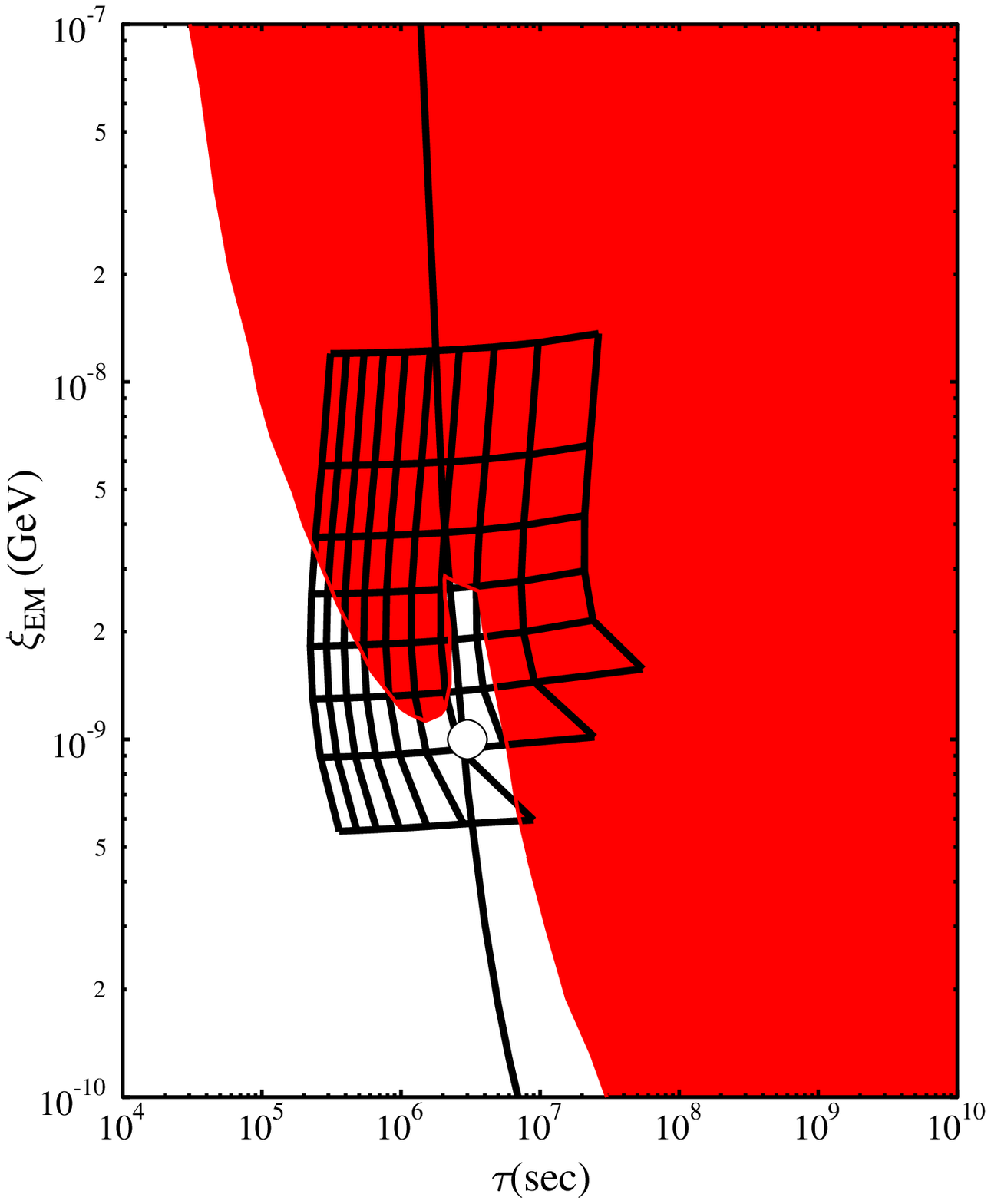}
\includegraphics[scale=0.25]{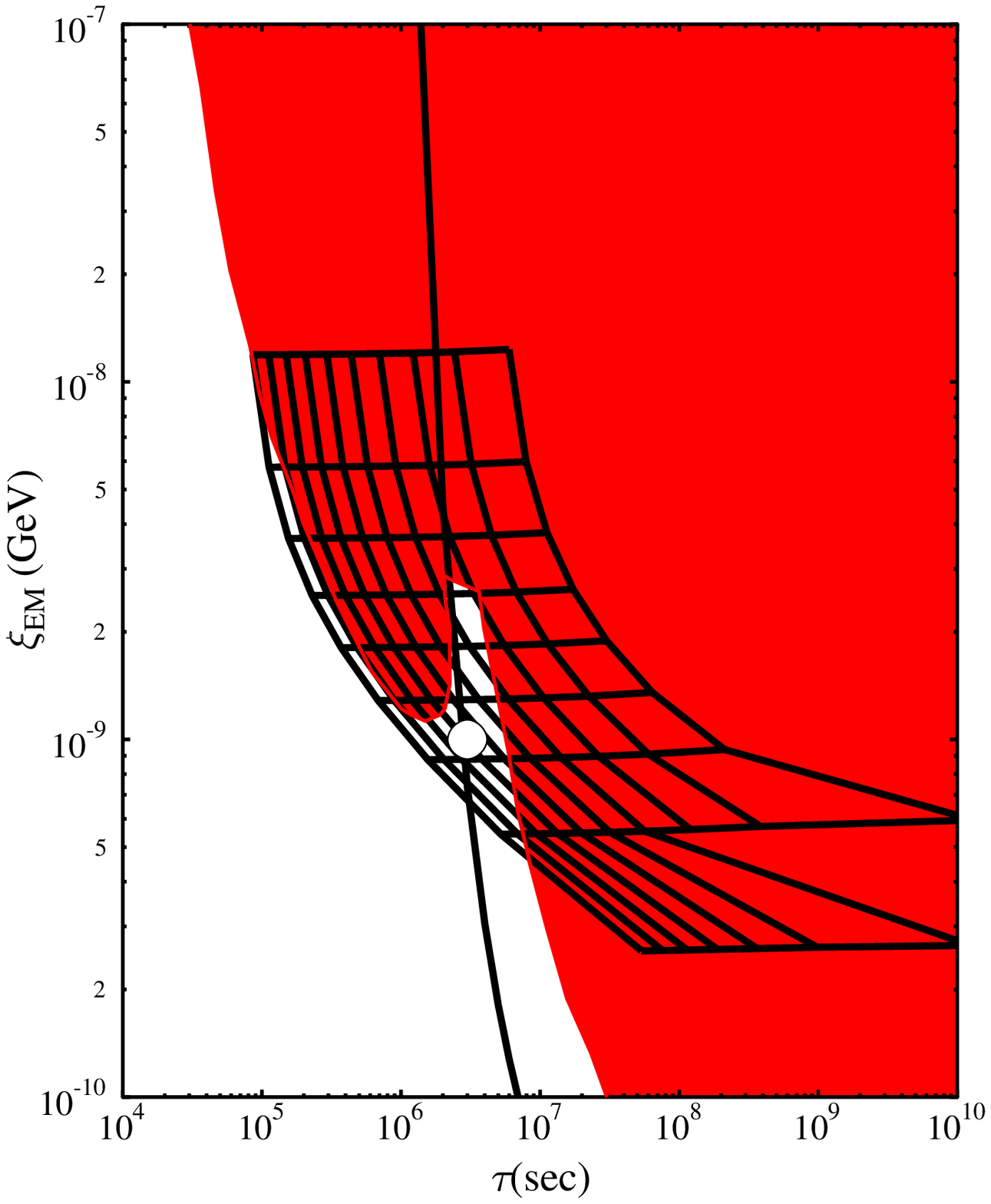}
\caption{\label{higgs}
Lifetime $\tau$ and energy release $\xi_{EM}$ in CP-even Higgsino decay for
$-\mu = 300 GeV, 400 GeV, \ldots, 1.2 TeV$ (right to left)
and $x_Q=0.1, 0.2, \ldots, 0.8$ (top to bottom) in the left panel and
in CP-odd Higgsino decay for $-\mu =600 GeV, 800 GeV, \ldots, 2.4 TeV$
(right to left) and $x_Q=0.1, 0.2, \ldots, 0.9$ (top to bottom).
We take $m_{h^0}=115 GeV$ and $c=1$.
For other comments, see Fig. \ref{bino}.
}
\end{figure}

We now study the possibility whether Quintessino can form 
the dark matter (DM). 
A similar mechanism for gravitino and graviton DM
produced in the weakly interacting massive particle (WIMP) 
decays has been studied in Ref. \cite{feng}.
For the process $\tilde{B}\to\tilde{Q}\gamma$, taking place
long after BBN, 
the total energy release in photons is severely constrained
by the BBN observation\cite{bbn}.
The CMB data also constrain the energy injection in form
of photons in order not to distort the Planckian CMB spectrum \cite{cmb}.
The electromagnetic energy released from the Bino decay can be written as
\begin{equation}
\xi_{EM}=\epsilon_{EM}N_{\tilde{B}}
=1.5\times 10^{-9} GeV \frac{1-x^2}{x}\ ,
\end{equation}
where $\epsilon_{EM}$ is the initial electromagnetic energy
released in each Bino decay and
$N_{\tilde{B}}=n_{\tilde{B}}/n_\gamma^{BG}$ is the number
density of Binos
normalized to the number density of the background photons.
Demanding Quintessino give the correct DM density, $\Omega_{DM}=0.23$,
$N_{\tilde{B}}$ is
given by\cite{feng} $N_{\tilde{B}}=3.0\times 10^{-12}
\left[ \frac{TeV}{m_{\tilde{Q}}} \right]
\left[ \frac{\Omega_{DM}}{0.23} \right] $.
Taking $c=3\times 10^{-2}$, the lifetime and the energy
release are generally greater than $10^7$ sec and $10^{10}$ GeV respectively,
which are, unfortunately, excluded by
the BBN\cite{bbn} and CMB\cite{cmb} constraints (see the excluded region
in the Figs. \ref{bino} and \ref{higgs}). Note that the axino dark 
matter particles are mainly produced by the anomalous 
interactions\cite{axino} similar
to Eq. (\ref{qphoton}). 
The argument and calculations show the differences 
between the axino and Quintessino.

To evade the BBN and CMB constraints, we introduce new interactions
between Quintessence and ordinary matter.
The shift symmetry, $Q \rightarrow Q + C$, implies that the interactions of
the Quintessence with matter should involve derivatives. 
In terms of an effective Lagrangian there are generally two classes of 
operators at dimension 5, one with fermions $f$ \cite{li} and the other one 
with Higgs boson $H$ of the standard electroweak theory 
\begin{eqnarray}
\label{qff}
\mathcal{L}_{Qff}&=&\frac{ 1 }{\Lambda}\partial_\mu Q 
( c_{ij}^R \bar{f_i}_R\gamma^\mu f_{j R}
+ c_{ij}^L \bar{f_i}_L\gamma^\mu f_{j L}) \ ,\\
\label{qhh}
\mathcal{L}_{QHH}&=&\frac{c_H}{\Lambda} i\partial_\mu Q \left(
H^\dagger D^\mu H - (D^\mu H)^\dagger H
\right)\ ,
\end{eqnarray}
where $\Lambda$ represents the cutoff energy scale and 
$D^\mu$ is the gauge covariant derivative.
Several constraints are set on the cutoff scale $\Lambda$.
First, since the Quintessence is very light, the coupling in the forms 
above will lead to an energy-loss channel for stars.
The cutoff is bounded below, $\Lambda \gtrsim 2\times 10^9 GeV$,   
in order not to lead conflict with
the observational limits on the stellar-evolution time scale\cite{revdicus}. 
The SN 1987A observation also constrains this ``invisible channel''
and leads to $\Lambda \gtrsim 6\times 10^9 GeV$\cite{ellis}.
The interactions in Eq. (\ref{qff}) 
also induce lepton flavor changing decay 
$ \mu \rightarrow e + Q$ with the branching ratio given
by Br$(\mu\to e Q)= \frac{3\pi^2}{\Lambda^2}\frac{1}{(m_\mu G_F)^2}$
for $c_L^{e\mu}=c_R^{e\mu}=1$.
The familon search experiments set the bound on the cutoff scale
as $\Lambda \gtrsim 4\times 10^9 GeV$\cite{bryman}. 
The operator in Eq. (\ref{qhh}) 
gives rise to a mixing between the Quintessence and 
the gauge boson $Z_{\mu}$, which induces an effective coupling of the 
Quintessence to the light fermions\cite{majo}. The astrophysical 
experiments put a limit $\Lambda \gtrsim 3\times 10^9 GeV$.
In a word, the present astrophysical and laboratory 
experimental limit on the
energy scale $\Lambda$ of an axion-like pseudoscalar coupling
with matter is around $ 
10^{10} GeV$\cite{pdg}. 

We supersymmetrize the interactions above by introducing
the gauge and supersymmetric invariant Lagrangian
\begin{equation}
\label{susylag}
\mathcal{L}= \frac{c}{\Lambda}
\hat{Q} \Phi^\dagger e^{2gV} \Phi|_{\theta\theta\bar{\theta}\bar{\theta}} + h.c.\ ,
\end{equation}
where $\hat{Q}=(\sigma_q+ i Q) +\sqrt{2} \theta \tilde{Q} +\theta\theta F$ is
the chiral superfield containing Quintessence $Q$ and its fermionic
partner $\tilde{Q}$, $\Phi$ is any matter superfield in the 
MSSM and $V$ is the vector superfield. We notice that this Lagrangian possesses
the shift symmetry, i.e., $\hat{Q}\to \hat{Q} + i\Lambda C $. 
When expressing it in terms of the component fields, we obtain
the needed couplings in Eqs. (\ref{qff}) and (\ref{qhh}).

Taking $\Phi$ in Eq. (\ref{susylag}) to be the Higgs superfield,
the Bino can decay via a new channel,
$\tilde{B}\to h^0\tilde{Q}$, with $h^0$ the lightest CP-even Higgs boson
and the relevant coupling given by
\begin{equation}
\label{bqh}
\mathcal{L}_{\tilde{B}h\tilde{Q}} =\frac{c}{\Lambda}
\frac{v}{\sqrt{2}} g'\cos(\alpha+\beta)
\bar{\tilde{B}}\tilde{Q}h^0\ ,
\end{equation}
where $v=246 GeV$ is the vacuum expectation value (VEV)
of the Higgs field, $g'$ is the gauge coupling of $U(1)_Y$,
$\tan\beta=v_2/v_1$ is the ratio between the two VEVs, and $\alpha$ is the 
mixing angle between the neutral Higgs bosons.
The decay width is given by
\begin{eqnarray}
\Gamma(\tilde{B}\to h\tilde{Q}) &&\approx
\left(\frac{c}{\Lambda}\right)^2\frac{\sin^2\theta_W}{8\pi}
M_Z^2 m_{\tilde{B}}\cdot \\
&&\left( (1-x_Q)^2-x_h^2 \right)^{\frac{1}{2}}
\left( (1+x_Q)^2-x_h^2 \right)^{\frac{3}{2}},\nonumber
\end{eqnarray}
where we have taken the limit of large $M_{A^0}$ and large $\tan\beta$,
with $x_Q=m_{\tilde{Q}}/m_{\tilde{B}}$ and $x_h=m_{h^0}/m_{\tilde{B}}$.

If the cutoff scale $\Lambda$ is near the present experimental
limit, for example, $\Lambda\sim 10^{12} GeV$, we have the decay time 
$\tau\sim 10^{-5}sec$ for $m_{\tilde{B}}=1TeV$, 
which means that the neutralino decays shortly after
it freezes out, and much before the BBN. 
In this case, the constraint from BBN is quite weak.
The Quintessino, if still relativistic, will 
contribute to the energy density and change the expanding rate of the
Universe during BBN. This energy contribution, in general, can be 
expressed in terms of the effective
number of extra generations of neutrinos, defined by $\delta
N_{\nu}\equiv \rho_{\tilde{Q}}/\rho_{\nu}$, where
$\rho_{\tilde{Q}}$ and $\rho_{\nu}$ are the energy density 
of Quintessino and one species of
neutrino, respectively.
In order not to affect the Universe's expansion too
much during BBN, $\delta N_{\nu}$ is constrained to be less than 
$0.2-1$\cite{jre}. 
For $\Lambda$ around $10^{12}$ GeV and $m_{\tilde B} = 1$TeV, 
we obtain $\delta N_{\nu}\sim 10^{-5}$, which is much below the limit.

\begin{figure}
\includegraphics[scale=0.3]{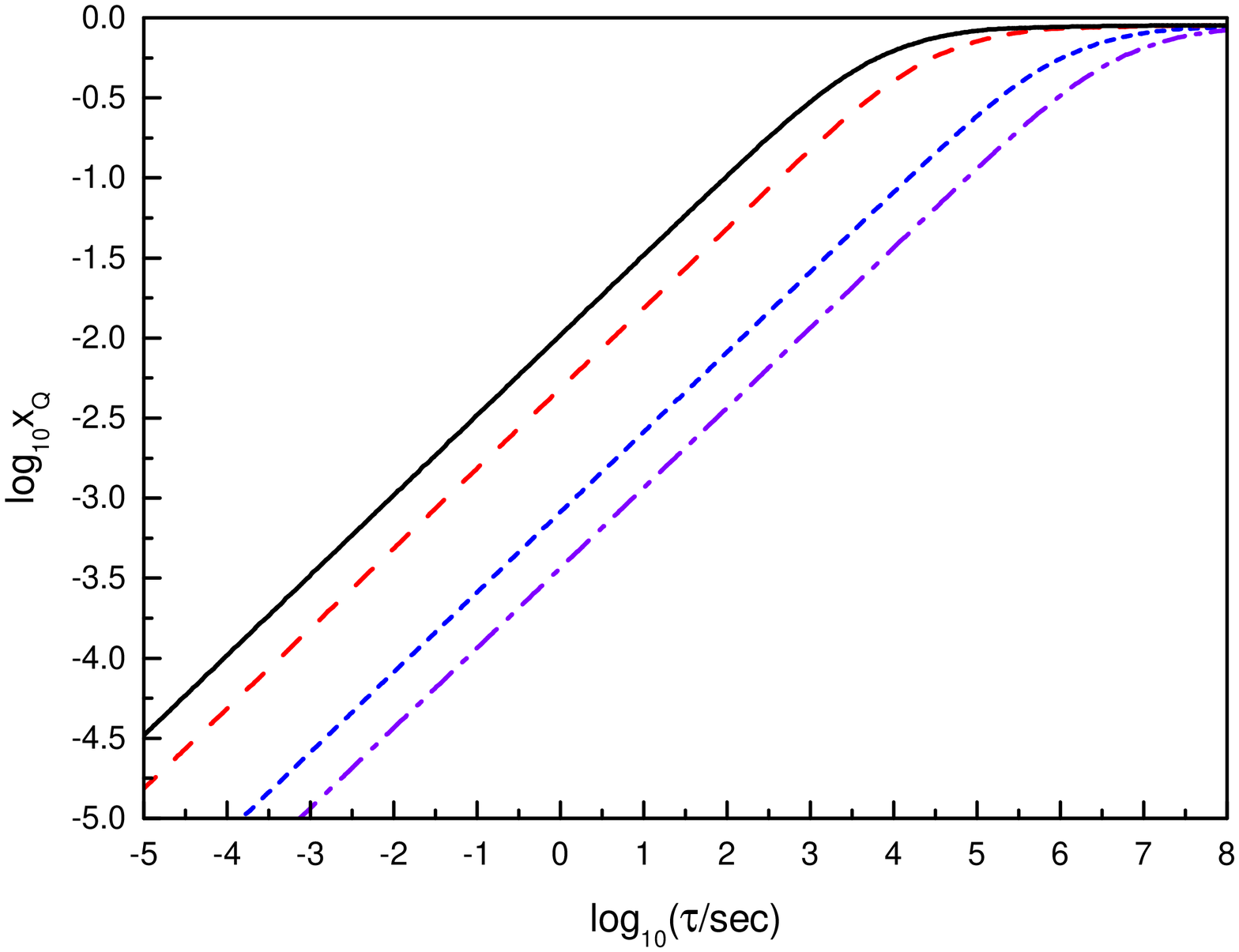}
\caption{\label{freestream}
The contours of the free streaming scale length 
$\lambda_{FS}=0.05$ Mpc (solid line), $0.1$ Mpc (dashed
line), $0.5$ Mpc (short dashed line), and $1.0$ Mpc (dashed-dotted line)
as function of $(x_Q,\tau)$ for the two body decay $\chi_1^0
\to \tilde{Q} h^0$. We fix $x_h=0.1$.} 
\end{figure}

The property of Quintessino dark matter produced non-thermally 
may be characterized by the
comoving free streaming scale, $\lambda_{FS}$, which represents a
quantity of crucial relevance to the formation of the large scale
cosmic structure. The density fluctuations on scales less than
$\lambda_{FS}$ would be severely suppressed.
We calculate the free streaming scale of Quintessino dark matter
as\cite{borganilin}
\begin{equation}
\lambda_{FS}=\int^{t_{EQ}}_{\tau}\frac{v(t')}{a(t')}dt'
\simeq  143.2  ~{\rm Mpc}~
 \ln \left(\frac{W+\sqrt{1+W^2}}{W}\right) ~ ,
\end{equation}
with $W=2 x_Q/\sqrt{((1+x_Q)^2-x_h^2)((1-x_Q)^2-x_h^2)y}$.
We have integrated the red-shifted velocity of Quintessino, $v/a$,
from the time when it is produced, $\tau$, to $t_{EQ}$ when the matter 
begins to dominate the Universe and $y\equiv \tau/t_{EQ}$. 
In Fig. \ref{freestream}, we show the curves of
constant $\lambda_{FS}$ in the $(\log_{10}(x_Q),~\log_{10}(\tau/sec))$
space. We can see that the free-streaming scale
of Quintessino, taking $\tau\sim 10^{-5}$ sec and
$x_Q\sim 10^{-3}$, is much shorter
than 0.1 Mpc, which means it serves as 
a good candidate for cold dark matter.

As the cutoff scale gets higher, the constraints from BBN and CMB
will come into action.  In the following, we will take
$\Lambda=M_{pl}$ for a detailed discussion.

For $\Lambda=M_{pl}$, the neutralino decays long after BBN, so we 
have to study the constraints from BBN and
CMB on the electromagnetic energy release in the 
decay. In general, the electromagnetic energies are released 
through the Higgs cascade decays.
The $h^0$ decays dominantly into $b\bar{b}$ and
$\tau\bar{\tau}$. We have calculated the averaged electromagnetic energy
released in the Higgs boson decay using the Jetset7.4 
Monte Carlo event generator package\cite{jetset} in the
decoupling limit.
We get $\epsilon_{EM} \approx 0.8E_{h^0}$, ignoring possible electromagnetic
energy carried by neutrinos via the process, such as,
$\nu\nu\to e^+e^-$. Since the present constraint on
the hadronic energy release is only sensitive to the decay time
$\tau \lesssim 10^4$ sec\cite{had}, we will not consider the constraint
from the hadronic energy release. In Fig. \ref{bino}, we plot the
decay lifetime and electromagnetic energy release for a range
of ($m_{\tilde{B}}$, $x_Q$). 
We notice that some parameter space is excluded by the BBN and CMB constraint,
but the region with $m_{\tilde{B}}\sim 400 GeV - 1.2 TeV$ and
$x_Q\sim 0.4 - 0.8$ remains viable. The best fit point
$(\tau, \xi_{EM})=(3\times 10^6 sec, 10^{-9} GeV)$ is marked in the figure,
which corresponds to such
a point that $^7$Li is destroyed to the level of
the present observation while keeping the concordance between 
CMB and BBN determination of the baryometer $\eta$ from other light elements.
The $^7$Li underabundance may
be an evidence supporting the late time WIMPs decay\cite{feng}.
In our scenario, this point corresponds to taking 
$m_{\tilde{B}}\approx 500 GeV$ and $x_Q\approx 0.7$.

Finally, we consider the case that the NLSP is Higgsino.
In the hyperbolic branch 
of the mSUGRA parameter space,
the lightest Higgsino is the LSP and gives correct
relic density even the neutralino mass is large,
due to the rapid coannihilation processes\cite{focus}.
In this case, for $\mu >0$ and $c=1$, the relevant coupling is 
given by
\begin{equation}
\mathcal{L}_{\tilde{H}\tilde{Q}h} = 
-\frac{1}{2}\frac{\mu}{M_{pl}}\left[C_S 
\bar{\tilde{H}}_S \tilde{Q}
-i C_A 
\bar{\tilde{H}}_A \gamma^5 \tilde{Q}\right] h^0\ ,
\end{equation}
with $C_{S,A}=\sin\alpha\mp \cos\alpha$ and
 $\tilde{H}_{S,A}\approx (\tilde{H}^1_1\mp \tilde{H}^2_2)/\sqrt{2}$
being the CP-even(odd) Higgsino. For $\mu<0$, 
$C_S$ and $C_A$ interchange, which become equal in the limit 
of large $\tan\beta$. 
In Fig. \ref{higgs}, we plot
the lifetime and energy release of the CP-even (left-panel) and
CP-odd (right-panel) Higgsino decay for a range of
($\mu$, $x_Q$) for $\mu < 0$. 
For the CP-even Higgsino we get the best fit point
by taking $\mu\approx -600 GeV$ and $x_Q\approx 0.7$.
For the CP-odd Higgsino
this point is reached when $\mu\approx -2 TeV$ and $x_Q\approx 0.7$.

From Fig. \ref{freestream} we can see that
the free-streaming scales
are in the range $\lambda_{FS} \sim 0.1-1.0$ Mpc
for $\tau\sim 10^6$ sec and $x_Q\sim 0.4-0.8$. 
So, for this
parameter space, the Quintessino has the properties similar to that of
warm dark matter (WDM), which has been proposed to resolve the
difficulties with the conventional WIMPs CDM model on subgalactic
scales \cite{borganilin,colin}. However, the WDM scenario
is constrained severely by the evidences for early galaxy and star
formation. The high optical depth of reionization found by WMAP 
data\cite{wmap} implies an early star formation at $z>10$. If this result is
confirmed, there would be no room for the presence of significant
WDM\cite{ostriker}. 
We leave the detailed investigation of the effects of
Quintessino DM and superWIMP\cite{feng} DM 
on the large scale structure and CMB for the future study \cite{bi}. 

In summary, we have studied the cosmological phenomenologies of models 
with supersymmetric interaction of Quintessence and matter.
We proposed new interactions between
Quintessence and matter, and examined the constraints by
the current astrophysical and laboratory experimental data.
We then studied their implications in Quintessino dark matter.
Our results show that Quintessino can be a good candidate for CDM or
WDM through the late-time decay of the NLSP of ordinary superparticle,
which can be the Bino or Higgsino-like neutralino. 
In our model, the present acceleration of the Universe is driven by the 
dynamics of Quintessence 
and, at the same time, the superpartner of Quintessence, the Quintessino, 
makes up the dark matter of the Universe\footnote{Note that the axion
dynamics will not be able to drive the Universe acceleration. Furthermore,
for axion and axino they both are dark matter particles.}. 

\begin{acknowledgments}
We would like to thank B. Feng, J. Feng, Y.N. Gao, 
P. Gondolo and R. Mohapatra,  for comments and discussions.
This work is supported in part by the NSF of China 
under the grant No. 10105004, 19925523, 10047004 
and also by the Ministry of Science and Technology of China under 
grant No. NKBRSF G19990754.
\end{acknowledgments}

\end{document}